\documentclass[showpacs,twocolumn,amsmath,aps]{revtex4}
\usepackage{graphicx,color}
\usepackage{bm}
\usepackage[hypertex]{hyperref}

\newcommand{\be}{\begin{equation}}
\newcommand{\ee}{\end{equation}}
\newcommand{\bea}{\begin{eqnarray}}
\newcommand{\eea}{\end{eqnarray}}
\newcommand{\bsube}{\begin{subequations}}
\newcommand{\esube}{\end{subequations}}

\newcommand{\Eq}[1]{Eq.\,(\ref{#1})}

\newcommand{\dg}{\dagger}
\newcommand{\la}{\langle}
\newcommand{\ra}{\rangle}

\newcommand{\nl}{\nonumber \\}

\usepackage{amsfonts}
\usepackage{amsmath}
\usepackage{amssymb}
\usepackage{graphicx}
\usepackage{type1cm}        
\usepackage{times}          
\usepackage{bm}
\usepackage{color}                      

\newcommand{\beq}{\begin{equation}}
\newcommand{\eeq}{\end{equation}}
\newcommand{\beqn}{\begin{eqnarray}}
\newcommand{\eeqn}{\end{eqnarray}}
\newcommand{\bsub}{\begin{subequations}}
\newcommand{\esub}{\end{subequations}}

\newcommand{\ket}[1]{{\left| #1 \right\rangle }}
\newcommand{\bra}[1]{{\left\langle #1 \right|}}

\def\Im{{\rm Im}}
\def\Re{{\rm Re}}

\begin{document}


\title{ Quantum Bayesian rule for weak measurements of
        qubits in superconducting circuit QED }

\author{Peiyue Wang}
\affiliation{Department of Physics, Beijing Normal University,
Beijing 100875, China}
\author{Lupei Qin}
\affiliation{Department of Physics, Beijing Normal University,
Beijing 100875, China}
\author{Xin-Qi Li }
\email{lixinqi@bnu.edu.cn}
\affiliation{Department of Physics, Beijing Normal University,
Beijing 100875, China}

\begin{abstract}

Compared with the quantum trajectory equation (QTE),
the quantum Bayesian approach has the advantage of being
more efficient to infer quantum state under monitoring,
based on the integrated output of measurement.
For weak measurement of qubits in circuit quantum electrodynamics (cQED),
properly accounting for the measurement backaction effects
within the Bayesian framework
is an important problem of current interest.
Elegant work towards this task was carried out by Korotkov
in ``bad-cavity" and weak-response limits (arXiv:1111.4016).
In the present work, based on insights from the cavity-field
states (dynamics) and the help of an effective QTE, we generalize
the results of arXiv:1111.4016 to more general system parameters.
The obtained Bayesian rule is in full agreement
with Korotkov's result in limiting cases and as well holds
satisfactory accuracy in non-limiting cases
in comparison with the QTE simulations.
We expect the proposed Bayesian rule to be useful
for future cQED measurement and control experiments.
\end{abstract}

\date{\today}
\maketitle


\section{Introduction}

The circuit quantum
electrodynamics (cQED) setup \cite{Bla04,Sch04,Mooij04}
is widely regarded as a promising solid-state architecture
for quantum computing and quantum information processing.
In the early stage, this setup is also an excellent
platform for quantum measurement and control studies
\cite{FN05,Sch08,Pala10,DiCa13,Hof11,Mar11,Dev13,Sid13,
Mil05,Li10,Li11,Goa14,Sid12,DiCa12,Dev13a}.
Particular examples include:
the experimental test of the Leggett-Garg inequality \cite{Pala10},
the measurement of weak values \cite{DiCa13},
the quantum back-action effect of weak measurements \cite{Dev13,Sid13},
and quantum feedback control experiments \cite{Sid12,DiCa12,Dev13a}.
In these studies, quantum measurements play a central role,
i.e., in dispersive regime \cite{Gam08,Gam09},
a dyne-type quadrature measurement of the cavity field
can reveal the qubit state information
\cite{Pala10,DiCa13,Hof11,Mar11,Dev13,Sid12,DiCa12,Dev13a}.

In this context, rather than strong projective measurement,
more interesting is the type of weak measurement
\cite{Har06,Car93,WM09} whose experimental realization
is an extremely attracting subject \cite{Dev13,Sid13}.
In particular, this type of monitoring on quantum state
is an essential prerequisite for measurement-based
feedback control of quantum systems \cite{WM09}.
For continuous weak measurements most popular
is the quantum trajectory theory \cite{WM09}, as broadly
applied in quantum optics and quantum control problems.
The quantum trajectory theory can also address the solid-state
charge qubit measurements by mesoscopic quantum-point-contact
and single-electron-transistor detectors \cite{Sch01,Goan01}.
For this setup, an equivalent scheme known as quantum Bayesian approach
was proposed \cite{Kor99} and exploited for applications.
For cQED, which is analogous to the conventional optical
cavity-QED, the quantum trajectory approach seems
the most natural choice \cite{Mil05,Li10,Li11,Gam08,Gam09}.
Despite this, in a recent study \cite{Kor11},
Korotkov developed a promising quantum Bayesian approach
in the ``bad-cavity" and weak-response limits.
Owing to its competitive efficiency and advantage
of accounting for realistic imperfections,
this approach has been employed in recent
experiments on quantum measurement \cite{Sid13}
and feedback control \cite{Sid12}.

Construction of the quantum Bayesian approach
is largely based on the classical Bayes formula.
For the diagonal elements of the qubit,
the Bayes formula works perfectly;
however, it does not work for the off-diagonal elements.
One proceeds then by a {\it purity} consideration \cite{Kor99},
together with some additional physical insights \cite{Kor11}.
In this work, rather than using such type of considerations,
we would like to fulfill a similar task by employing
the quantum trajectory equation (QTE) approach.
In order to gain necessary insights,
our analysis will pay particular attention to the nature
of the cavity field under continuous quadrature monitoring.
This treatment permits us to avoid the bad-cavity
and weak-response assumptions \cite{Kor11}, making thus the obtained
quantum Bayesian rule applicable to more general setup parameters.

The paper is organized as follows. We begin in Sec.\ II
with a brief description of cQED and the optical QTE,
before identifying the nature of the cavity state
conditioned on weak measurements in Sec.\ III.
We then construct in Sec.\ IV the quantum Bayesian rule
for single-quadrature measurements
and in Sec.\ V for two-quadrature measurements.
In Sec.\ VI, numerical results and comparisons are presented,
for both the limiting and non-limiting cases.
Finally, we summarize the work with remarks in Sec.\ VII.
In addition, two Appendices are provided, for
the analytic solution of the cavity field and an equivalence
proof of two expressions of the purity degradation factor.

\section{Model and Quantum Trajectory Equation}

Let us consider the simplest cQED setup with only one
superconducting qubit in the resonator cavity \cite{Bla04}.
In this setup,
the central section of superconducting coplanar waveguide
plays the role of an optical cavity,
and the superconducting qubit the role of an (artificial) atom.
The superconducting qubit is coupled to the one-dimensional
transmission line (1DTL) cavity which acts as a simple harmonic oscillator.
Therefore the qubit, the 1DTL cavity, and their mutual coupling
can be well described by the Jaynes-Cummings Hamiltonian.
Moreover, we consider the setup in a dispersive regime
\cite{Bla04,Sch04,Mooij04}, i.e., with the detuning between the
cavity frequency ($\omega_r$) and qubit energy ($\omega_q$),
$\Delta=\omega_r-\omega_q$, much larger than the coupling strength $g$.
In this limit and in the rotating frame
with the microwave driving frequency $\omega_m$,
the system can be described by an effective Hamiltonian \cite{Bla04,Sch04}
\begin{equation}\label{H_eff}
H_{\rm eff}=\Delta_r a^{\dagger}a
  +\frac{\widetilde{\omega}_q}{2}\sigma_z
  +\chi a^{\dagger}a\sigma_z
  +\left(\epsilon_m^\ast a+\epsilon_m a^{\dagger}\right) \;,
\end{equation}
where $\Delta_r=\omega_r-\omega_m$
(for resonant drive, $\Delta_r=0$),
and $\widetilde{\omega}_q=\omega_q+\chi$
with $\chi=g^2/\Delta$ being a dispersive shift
to the qubit energy and cavity frequency.
In \Eq{H_eff}, $a^{\dagger}$ ($a$) and $\sigma_z$ are respectively
the creation (annihilation) operator of cavity photon
and the quasi-spin operator (Pauli matrix) for the qubit.
$\epsilon_m$ is the microwave drive amplitude to the cavity.

For measurements, in this work we will first consider the
single quadrature ($I_{\varphi}$) measurement in detail,
then convert the obtained results
to the $(I,Q)$ two-quadrature measurement.
For the single quadrature homodyne measurement, one actually measures
$\hat{I}_\varphi= \frac{1}{2}(ae^{-i\varphi}+a^\dag e^{i\varphi})$,
where $\varphi$ is the local oscillator (LO) phase \cite{WM09}.
The measurement output can be expressed as
\bea\label{Iphi}
I_{\varphi}(t)=\sqrt{\kappa}\langle
ae^{-i\varphi}+a^\dag e^{i\varphi}  \rangle_{\varrho(t)}  +\xi(t) \;,
\eea
where $\kappa$ is the damping rate of the cavity photon
and $\xi(t)$ a Gaussian white noise
originating from the stochastic quantum-jump,
which satisfies the ensemble-average
properties of $E[\xi(t)]=0$ and $E[\xi(t)\xi(t')]=\delta(t-t')$.
The quantum average $\langle \cdots \rangle_{\varrho(t)}$
is defined by $\langle \cdots \rangle_{\varrho(t)}
={\rm Tr}[(\cdots)\varrho(t)]$, with $\varrho(t)$
the qubit-cavity conditional state
given by the quantum trajectory equation (QTE) \cite{WM09}:
\begin{eqnarray}\label{QTE}
 \dot{\varrho} = -i[H_{\rm eff},\varrho]
    +\kappa\mathcal{D}[a]\varrho
   +\sqrt{\kappa}\mathcal{H}[ae^{-i\varphi}] \varrho \xi(t) \;,
\end{eqnarray}
where $\mathcal{D}[a]\varrho=a\varrho a^{\dagger}
-\frac{1}{2}\{a^{\dagger}a,\varrho \}$ and
$\mathcal{H}[\bullet]\varrho = (\bullet)\varrho+\varrho (\bullet)^{\dg}
-\mathrm{Tr}\{[(\bullet)+(\bullet)^{\dg}]\varrho\}\varrho$.


\section{Cavity State Conditioned on Weak Measurements}

To get necessary insights for constructing
a quantum Bayesian rule for the qubit state,
it is crucial to identify the nature of the cavity state
conditioned on the quadrature outcomes.
First, we notice that,
for a specific qubit state $\ket{g}$ or $\ket{e}$, the interplay
of measurement drive and cavity loss would lead to the formation
of a coherent state $\ket{\alpha_g(t)}$ or $\ket{\alpha_e(t)}$
for the cavity field,
with $\alpha_{g(e)}(t)$ determined by the following equations:
\begin{eqnarray}\label{alphat}
 && \dot{\alpha_e}(t)=-i\epsilon_m
 -i(\Delta_r+\chi)\alpha_e(t)-\kappa\alpha_e(t)/2 \;,   \nonumber\\
 && \dot{\alpha_g}(t)=-i\epsilon_m
 -i(\Delta_r-\chi)\alpha_g(t)-\kappa\alpha_g(t)/2 \;.
 \end{eqnarray}
Notice also that this result is associated with an ensemble average
over the stochastic leakage of photons.
In the stationary limit, the coherent-state parameter reads
$\alpha_{g(e)}=-i\epsilon_m/(i\Delta_{r,g(e)}+\kappa/2)$,
where $\Delta_{r,g(e)}=(\omega_r-\omega_m)\mp \chi$.
The transient solution is also available
(but with a lengthy expression, see Appendix A).

As a heuristic discussion for measurement principle,
let us first consider a simpler model of
qubit measurement by another two-state meter (e.g., a spin),
which are prepared in an entangled initial state
$c_g\ket{g}\otimes \ket{\uparrow} + c_e \ket{e}\otimes \ket{\downarrow}$.
Here $\ket{\uparrow}$ and $\ket{\downarrow}$ are the meter basis states
(in the $\sigma_z$ representation).
Then, let us consider a projective measurement on the spin
in a different (e.g., $x$) direction.
A specific result, for instance ``$+1$" in the $x$-direction,
would project the joint state onto
$(c_g d_{\uparrow} \ket{g} + c_e d_{\downarrow} \ket{e})
\otimes \ket{\uparrow}_x$, where
$d_{\uparrow}=~ _x\bra{\uparrow}\uparrow\ra$ and
$d_{\downarrow}=~ _x\bra{\uparrow}\downarrow\ra$.
Since the measurement basis
$\ket{\uparrow}_x$ ($\ket{\downarrow}_x$)
is not parallel to $\ket{\uparrow}$ ($\ket{\downarrow}$),
the strong projective measurement on the meter does not
collapse the qubit state onto $\ket{g}$ or $\ket{e}$.
To the qubit state, this falls into the category of weak measurements.

Now consider the qubit measurement in cQED.
The qubit-cavity state is initially prepared as
$\ket{\Psi(0)}=(c_g \ket{g} + c_e \ket{e})\otimes\ket{\alpha_0}$
($\alpha_0=0$ if the cavity field is the vacuum).
If one is faithfully tracking
the emitted photon by continuous homodyne measurement,
the subsequent time-dependent state can be expressed as
\begin{eqnarray}\label{state-1}
\ket{\Psi(t)}=
c_g(t) \ket{g}\otimes\ket{\tilde{\alpha}_g(t)}
+ c_e(t) \ket{e}\otimes\ket{\tilde{\alpha}_e(t)} \;.
\end{eqnarray}
Here, instead of $\ket{\alpha}_{g(e)}(t)$,
we denote the respective cavity state as
$\ket{\tilde{\alpha}_{g(e)}(t)}$,
indicating a lack of ensemble average.
Based on the quantum measurement theory \cite{WM09},
this state is a {\it stochastic} and quantum {\it pure} state.
That is, both the superposition coefficients $c_{g(e)}(t)$
and the cavity states $\ket{\tilde{\alpha}_{g(e)}(t)}$
are stochastic, depending on the random outputs of measurement.
Now, consider a further weak measurement on this state,
with an {\it integrated} quadrature output $I_m$
over the time interval $(t, t+\tau)$.
One may imagine that the joint state of
the qubit-plus-cavity could be expressed as
\bea\label{state-2}
\ket{\Psi(t+\tau)}=
\left( d_g c_g \ket{g} + d_e c_e \ket{e}\right)
 \otimes\ket{\psi_{m}(t+\tau)} \;,
\eea
where the state update factors are given by
$d_{g(e)}=\la \psi_{m}(t+\tau)|\tilde{\alpha}_{g(e)}(t)\ra$.
The essential feature of this result is that the cavity field
has collapsed onto a unique eigenstate $\ket{\psi_{m}}$
of the quadrature operator, conditioned on the measurement record $I_m$.
However, we will show that this is not true.

As an explicit demonstration,
we performed a direct simulation of
\Eq{QTE} for continuous homodyne measurements, starting
with a superposition qubit state and a vacuum cavity state.
Based on the {\it conditional} joint qubit-plus-cavity state
$\varrho(t)$, the cavity states $\ket{\tilde{\alpha}_g(t)}$
and $\ket{\tilde{\alpha}_e(t)}$
in \Eq{state-1} can be extracted from $\varrho(t)$, respectively,
in terms of density matrix
$\varrho_{gg}(t)=\bra{g} \varrho(t) \ket{g}
/{\rm Tr}[\bra{g} \varrho(t) \ket{g}]$
and $\varrho_{ee}(t)=\bra{e} \varrho(t) \ket{e}
/{\rm Tr}[\bra{e} \varrho(t) \ket{e}]$.
Here ${\rm Tr}[\cdots]$ is over the cavity degrees of freedom.
Using a $Q$-function representation, in Fig.\ 1 we plot the difference
of $\varrho_{gg}(t)$ and $\varrho_{ee}(t)$ at a given time.
We find that this type of {\it coexistence} of {\it distinct} coherent
states will persist along the whole continuous weak measurement process.
This result indicates that the weak measurement
(with an outcome $I_m$) during $(t,t+\tau)$ does {\it not}
collapse the cavity fields $\ket{\tilde{\alpha}_{g(e)}(t)}$
onto any common eigenstate $\ket{\psi_{m}}$.
This differs substantially
from the simple two-state-meter example discussed earlier.

We may contrast the {\it state structure} revealed here
with the assumption in Ref.\ \cite{Dev13}.
In the Supplementary Materials (in Sec.\ 2.1.2, before Eq.\ (6)),
the following statement was made for the two-quadrature measurement:
the measurement process, with quadrature outcomes $(I_m,Q_m)$,
would project the cavity modes onto a unique eigenstate.
While the final result obtained there seems correct,
it might be of interest to further clarify their treatment.

\begin{figure}
\includegraphics[scale=0.55]{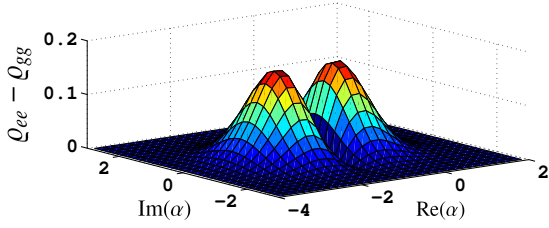}
\caption{ (Color online)
$Q$-function of the cavity field
during continuous weak measurements
(at a specific time $t=5 \kappa^{-1}$).
The definitions used here are
$\varrho_{ee}=\langle e|\varrho|e\rangle/
\mathrm{Tr}[\langle e|\varrho|e\rangle]$ and
$\varrho_{gg}=\langle g|\varrho|g\rangle/
\mathrm{Tr}[\langle g|\varrho|g\rangle]$,
with ${\rm Tr}[\cdots]$ over the cavity degrees of freedom.
Plotting the $Q$-function
$\la\alpha|(\varrho_{ee}-\varrho_{gg})|\alpha\ra/\pi$
reveals that the cavity field does not collapse onto
a unique eigenstate of the quadrature operator (i.e., the observable)
after experiencing a weak quadrature-measurement.
Parameters: $\Delta_r=0$, $\chi=0.1$,
$\epsilon_m=1.0$, $\kappa=2.0$, $\varphi=0$.     } 
\label{cavitystate}
\end{figure}

More careful inspection
reveals that the states $\varrho_{gg}(t)$
and $\varrho_{ee}(t)$ are very close to the coherent states
$|\alpha_g(t)\ra$ and $|\alpha_e(t)\ra$, respectively.
As we will see below, this identification can help us to
determine the {\it purity degradation factor} in the Bayesian rule.
Here, one may understand this essential result as follows.
Rather than a direct point-process detection for the
outgoing photon, the homodyne-type quadrature measurement
is relatively soft to the cavity field.
It does not drastically alter the number of cavity photons.
The continuous quadrature measurement is mainly updating
our knowledge about the superposition components, say,
the coefficients $c_g(t)$ and $c_e(t)$ in \Eq{state-1},
but not collapsing the cavity states
$\varrho_{gg}(t)$ and $\varrho_{ee}(t)$
onto a common eigenstate $\ket{\psi_{m}}$.
This is the so-called {\it informational} evolution \cite{Katz06,Kor07+07}.
However, around $\alpha_g(t)$ and $\alpha_e(t)$,
the cavity field {\it does} exist stochastic fluctuations,
which will result in a stochastic phase factor
to the qubit off-diagonal elements
-- as we will see in the following sections.

Finally, we remark that this cQED is an interesting
way to understand the puzzle of Schr\"odinger's cat.
That is, the cavity state $\ket{\tilde{\alpha}_{g(e)}(t)}$
corresponds to the macroscopic state (``dead" or ``alive") of the cat,
while the continuum of states outside the cavity
corresponds to the infinite number of microscopic states of the cat.
It is right these infinite number of microscopic degrees of freedom
that destroy the coherence of the superposed entangled state.

\section{Bayesian Rule for Single Quadrature Measurement}

The conditional evolution of the qubit-plus-cavity
under continuous measurements is well captured by \Eq{QTE},
However, a full simulation of this equation is time-consuming
and almost intractable in practice
(e.g., in the quantum feedback control experiment \cite{Sid12}).
More efficient method is using a quantum Bayesian rule
to update the qubit state, merely based on the measurement record
$I_m$ over certain finite time interval $t_m$.
For the sake of clarity, below we present our construction procedures
in order by three steps (addressed by three subsections).

\subsection{Bare Bayesian Rule}

To perform a Bayesian inference based on $I_m$,
which is defined by
$I_m = \frac{1}{t_m}\int^{t_m}_{0}dt I_{\varphi}(t)$
and collected in experiment
from the quadrature measurement records,
we need to know in advance the distribution $P_{g(e)}(I_m)$
associated with the qubit state $\ket{g(e)}$.
As given in Appendix A, a simple analysis shows
that the distribution is Gaussian:
\bea
P_{g(e)}(I_m)=\frac{1}{\sqrt{2\pi D}}
\exp[-(I_m-\bar{I}_{g(e)})^2/(2D)] \;,
\eea
with $D=1/t_m$ being the variance.
The average quadrature outcome, $\bar{I}_{g(e)}$, is given by
\bea
\bar{I}_{g(e)} = \frac{1}{t_m}\int^{t_m}_{0}dt
\,(2\sqrt{\kappa})\,{\rm Re}[\alpha_{g(e)}(t)e^{-i\varphi}] \;,
\eea
where $\alpha_{g(e)}(t)$ is the cavity field discussed
in Sec.\ III, with explicit solution presented in Appendix A.

With the knowledge of $P_{g}(I_m)$ and $P_{e}(I_m)$,
using the classical Bayes formula one can determine
$|c_{g}(t_m)|^2$ and $|c_{e}(t_m)|^2$, in \Eq{state-1}.
Obviously, they coincide with the diagonal elements
of the reduced density matrix of the qubit state.
Therefore, we have
\begin{eqnarray}\label{Bayes3a-1}
&&\rho_{gg}(t_m)=\rho_{gg}(0)\;P_g(I_m)/{\cal N} \;, \nonumber \\
&&\rho_{ee}(t_m)=\rho_{ee}(0)\;P_e(I_m)/{\cal N} \;,
\end{eqnarray}
where ${\cal N}=\rho_{gg}(0)P_g(I_m)+\rho_{ee}(0)P_e(I_m)$.
One can examine that \Eq{Bayes3a-1} is in full agreement
with the quantum trajectory equation simulation, as expected.

Regarding the off-diagonal element $\rho_{ge}(t_m)$,
the situation is subtle.
No classical rule applies here.
Following Ref.\ \cite{Kor99}, based on
purity consideration, we preliminarily have
\bea\label{off-1}
\tilde{\rho}_{ge}(t_m)
=\rho_{ge}(0)e^{-i\widetilde{\omega}_q t_m}
\sqrt{P_g(I_m)P_e(I_m)}/{\cal N},
\eea
to approximate $\rho_{ge}(t_m)$.
However, as shown in Fig.\ 2(a),
this result differs considerably from
the exact one from a direct simulation of \Eq{QTE}.

\subsection{Purity Degradation Factor}

For further corrections to $\tilde{\rho}_{ge}(t_m)$,
let us look back to \Eq{state-1}.
Based on the joint-state structure,
we propose to amend \Eq{off-1} by a purity degradation factor as
\bea\label{off-2}
\rho_{ge}(t_m)
=\tilde{\rho}_{ge}(t_m)~|\la \alpha_e(t_m) | \alpha_g(t_m) \ra| \; .
\eea
The physical meaning of this correction factor
is an account of the purity degradation,
after partly averaging an entangled state.
More specifically,
as shown in Fig.\ 1 and discussed earlier around it,
we know that the cavity state
evolves along $| \alpha_{g(e)}(t) \ra$,
with some tiny stochastic fluctuations.
Here, as a first step, we introduce
the {\it purity degradation factor},
$D(t_m) \equiv |\la \alpha_e(t_m) | \alpha_g(t_m) \ra|$,
to characterize the decrease of the qubit coherence
owing to averaging the cavity state,
while studying later the fluctuation effects on the qubit.
Using the property of coherent state,
an explicit result can be obtained as
\bea
D(t_m)&=&\exp \{  -1/2
\,[|\alpha_{e}(t_m)|^2+|\alpha_{g}(t_m)|^2]   \nl
&& ~~~~~~~ + \Re[\alpha_{e}(t_m)\alpha_{g}^*(t_m)] \}  \,.
\eea
Moreover, as proved in Appendix B, this result is precisely
equivalent to the following one by a more sophisticated
analysis based on the quantum trajectory equation:
\bea
D(t_m)=\exp\left\{-\int_0^{t_m} dt
\,[\Gamma_d(t)-\Gamma_m(t)/2]\right\} \,,
\eea
where $\Gamma_d(t)$ is a time-dependent overall decoherence
rate of the qubit caused by measurement, and $\Gamma_m(t)$
is the measurement (unraveling) rate.
This identification reveals a deep connection between the two
very different treatments, and provides an additional
evidence for the validity of the purity degradation factor.


\subsection{ Effects of Dynamic and
Stochastic Fluctuations of the Cavity Field }

Equation (\ref{state-1}) and Fig.\ 1 jointly indicate the qualitative
feature of the cavity state
and guide our construction for the quantum Bayesian rule.
However, in order to quantify the fluctuation effects of the
cavity field, more sophisticated skill is useful.
Let us return to the quantum trajectory equation.
Based on \Eq{QTE}, via a qubit-state-dependent displacement
transformation, it is possible to eliminate
the cavity degrees of freedom from this equation \cite{Gam08}.
Below we continue our correction to $\rho_{ge}(t_m)$
based on the achievement of this technique.
The transformed QTE reads \cite{Gam08}
\begin{eqnarray}\label{pQTE}
 &&  \dot{\rho} = -i\frac{\tilde{\omega}_q+B(t)}{2}[\sigma_z,\rho]
   +\frac{\Gamma_d(t)}{2}\mathcal{D}[\sigma_z]\rho       \nonumber \\
 &&  -\sqrt{\Gamma_{ci}(t)} \mathcal{M}[\sigma_z]\rho\xi(t)
   +i\frac{\sqrt{\Gamma_{ba}(t)}}{2}[\sigma_z,\rho]\xi(t)\,. \nl
 \end{eqnarray}
In this result, the effective magnetic field
$B(t)=2\chi\mathrm{Re}[\alpha_g(t)\alpha_e(t)^*]$,
describes a generalized ac-Stark shift
of the qubit energy, as a consequence of
dynamic fluctuations of the cavity field
owing to dispersive coupling to the qubit.
The superoperator is defined by
$\mathcal{M}[\sigma_z]\rho \equiv (\sigma_z\rho+\rho\sigma_z)/2
-\la \sigma_z\ra \rho$,
with $\la \sigma_z\ra={\rm Tr}[\sigma_z\rho]$,
an average over the reduced density matrix of qubit.
Additionally,
\begin{eqnarray}\label{GM3t}
  &&  \Gamma_d(t)=2\chi\mathrm{Im}[\alpha_g(t)\alpha_e(t)^*] \,,       \nonumber \\
  && \Gamma_{ci}(t) = \kappa|\beta(t)|^2\cos^2(\varphi-\theta_\beta) \,,  \nonumber \\
  &&  \Gamma_{ba}(t) = \kappa|\beta(t)|^2\sin^2(\varphi-\theta_\beta)\,,
\end{eqnarray}
characterize, respectively, the ensemble-average dephasing,
single measurement information-gain, and back-action rates.
Moreover, the sum of $\Gamma_{ci}$ and $\Gamma_{ba}$,
\bea
\Gamma_m(t)=\Gamma_{ci}(t)+\Gamma_{ba}(t)=\kappa|\beta(t)|^2 \;,
\eea
gives the measurement rate \cite{Gam08}.
In the above rates, we have used the definition
$\beta(t)=\alpha_e(t)-\alpha_g(t)
\equiv |\beta(t)|e^{i\theta_{\beta}}$.

Returning to \Eq{pQTE},
one may notice two unitary terms on the r.h.s.:
the first term, involving $B(t)$; and the last
with an effective stochastic field,
$-\sqrt{\Gamma_{ba}(t)}\xi(t)$.
These two terms properly characterize
the cavity-field-fluctuation effects on the qubit
and can be used to amend \Eq{off-2} further. We thus complete
our correction to the off-diagonal element as follows:
\begin{eqnarray}\label{Bayes3b-1}
&& \rho_{ge}(t_m)= \tilde{\rho}_{ge}(t_m)
     ~ | \la \alpha_e(t_m)| \alpha_g(t_m) \ra |  \nl
&&  ~~~~~~~~~~~~  \times   \exp\{-i[\Phi_1(t_m) + \Phi_2(t_m)]\} \;,
\end{eqnarray}
with the two additional phase factors
\begin{eqnarray}\label{Phi2a-1}
\Phi_1(t_m) &=& \int_0^{t_m} B(t) \;dt \,,  \nl
\Phi_2(t_m) &=& -\int_0^{t_m} \sqrt{\Gamma_{ba}(t)}\;
\widetilde{I}_{\varphi}(t) \;dt \,.
\end{eqnarray}
For $\Phi_2(t_m)$, $\widetilde{I}_{\varphi}(t)$ in the integrand 
is determined from the following considerations. 
First, within the ``polaron" transformation scheme, 
the output current can be reexpressed as \cite{Gam08}:
$ I_{\varphi}(t)
= -\sqrt{\Gamma_{ci}}\la \sigma_z \ra
+ \sqrt{\kappa}|\mu|\cos(\theta_{\mu}-\varphi)+\xi(t)$ , 
where $\mu=\alpha_e(t)+\alpha_g(t)\equiv |\mu|e^{i\theta_{\mu}}$.
Second, ensemble average of $\rho_{ge}(t_m)$
over the stochastic ``field" $\xi(t)$, or equivalently 
over the stochastic output current $I_{\varphi}(t)$,
should be consistent with the result by averaging \Eq{pQTE}.
Therefore, in the formal integrated solution
$\Phi_2(t_m) = -\int_0^{t_m} \sqrt{\Gamma_{ba}(t)}\;\xi(t) \;dt $, 
we need to replace ``$\xi(t)$" with $\widetilde{I}_{\varphi}(t)
= I_{\varphi}(t) - \sqrt{\kappa}|\mu|\cos(\theta_{\mu}-\varphi)
\equiv I_{\varphi}(t)-\bar{I}_{\varphi}(t)$.

Equations (\ref{Bayes3b-1}) and (\ref{Phi2a-1}),
together with (\ref{Bayes3a-1}) and (\ref{off-1}),
constitute the quantum Bayesian rule we propose
for qubit state under single-quadrature weak measurements.
To implement the proposed Bayesian rule in real experiments,
one can directly use the analytic solutions of
$\alpha_g(t)$ and $\alpha_e(t)$,
given in Appendix A,
and compute all the relevant quantities including
$| \la \alpha_e(t_m)| \alpha_g(t_m) \ra |$ and $\Phi_1(t_m)$.
Provided the necessary setup parameters are determined
(as discussed in further detail below), all these calculations 
can be fulfilled in advance. One can then update the qubit state. 
For further possible simplification, 
we propose here a scheme to approximate $\Phi_2$.
From Fig.\ 1 and the related discussion, we know that
$\Gamma_{ba}(t)$ is a slowly-varying function of time.
Thus, we propose
\begin{eqnarray}\label{Phi2}
\Phi_2(t_m) \simeq -\sqrt{\Gamma_{ba}(t_m)}\,[I_m-\bar{I}(t_m)]\,,
\end{eqnarray}
where 
\bea
\bar{I}(t_m)= \frac{1}{t_m}\int^{t_m}_{0} dt \bar{I}_{\varphi}(t)\,. 
\eea
As we will demonstrate, this approximation can work well
and could be useful in practice.

\section{ Bayesian Rule for Two-Quadrature Measurement }

In practice, rather than the $I_{\varphi}$
single-quadrature measurement,
the so-called $(I,Q)$ two-quadrature measurement
is another choice \cite{Dev13}.
Among the various realizations \cite{WM09},
it can be implemented as follows
(in terms of a quantum optics language, for convenience).
Using a beam-splitter, the outgoing microwave field
leaked from the cavity, $\sqrt{\kappa}a$,
is split into two branches with amplitudes
$a_1=\sqrt{\kappa/2}a$ and $a_2=i\sqrt{\kappa/2}a$.
Then, perform single-quadrature homodyne measurement on each branch
for $a_1$ and $a_2$, choosing the LO phases as
$\varphi=0$ and $\pi$, respectively.
One can prove that, for the cavity field, this realizes
an $I$-quadrature measurement in the first branch
and a $Q$-quadrature measurement in the second branch,
with measurement outputs described by
$I_m(t)=\sqrt{\kappa/2}
\la a +a^{\dagger}\ra_{\varrho(t)} + \xi_1(t)$, and
$Q_m(t)=\sqrt{\kappa/2}
\la -i a + i a^{\dagger} \ra_{\varrho(t)} + \xi_2(t)$.
Conditioned on this type of $(I,Q)$ two-quadrature measurements,
the evolution of the qubit-cavity joint state
follows the quantum trajectory equation:
\begin{eqnarray}\label{IQ-SME}
&&\dot{\varrho} =
 -i[H_{\rm eff},\varrho]
    +\kappa\mathcal{D}[a]\varrho           \nonumber \\
&& ~ +\sqrt{\frac{\kappa}{2}}\mathcal{H}[a]\varrho \,\xi_1(t)
   +\sqrt{\frac{\kappa}{2}}\mathcal{H}[-ia]\varrho \,\xi_2(t)\,.
\end{eqnarray}
To the entire qubit-plus-cavity state, the last two terms
fully unravel the measurement with no information loss.
Since only the output of the first branch (leading to the third term)
reveals the qubit state information, the information-gain rate
is just half of the optimal single quadrature measurement.
However, provided we keep track of the $Q$-quadrature output
in the second branch,
the qubit state can be maintained in high purity.
In a similar manner to the single-quadrature measurement,
applying the qubit-state dependent displacement transformation
to the cavity field within the two-quadrature measurement framework
also yields an effective quantum
trajectory equation for the qubit state alone:
\begin{eqnarray}\label{IQ-pQTE}
 \dot{\rho} &=&  -i\frac{\tilde{\omega}_q+B(t)}{2}[\sigma_z,\rho]
 +\frac{\Gamma_d(t)}{2}\mathcal{D}[\sigma_z]\,\rho         \nonumber \\
 && -\sqrt{\Gamma_m(t)/2}\,\mathcal{M}[\sigma_z]\rho \,\xi_1(t) \nl
 &&  +i\frac{\sqrt{\Gamma_m(t)/2}}{2}\,[\sigma_z,\rho]\,\xi_2(t) \;.
\end{eqnarray}
Here, $B(t)$ and $\Gamma_d(t)$ are the same
as in the single quadrature measurement,
see Eqs.\ (\ref{pQTE}) and (\ref{GM3t}).
Comparing \Eq{IQ-pQTE} with Eqs.\ (\ref{pQTE}) and (\ref{GM3t})
indicates that
the $I$-quadrature measurement of the first branch is associated with
$\Gamma_{ci}^{(1)}(t)=(\kappa/2)|\beta(t)|^2=\Gamma_m(t)/2$
and $\Gamma_{ba}^{(1)}(t)=0$;
while conversely, the $Q$-quadrature measurement of the second branch
is associated  with $\Gamma_{ci}^{(2)}(t)=0$
and $\Gamma_{ba}^{(2)}(t)=(\kappa/2)|\beta(t)|^2=\Gamma_m(t)/2$.

Following the same procedures as
for the single-quadrature measurement,
we can construct a quantum Bayesian
rule for the two-quadrature measurement.
First, the integrated output distribution
of the two-quadrature measurement is:
\begin{eqnarray}
P_{g(e)}(I_m,Q_m)&=& \left( \frac{1}{2\pi D} \right)
\exp[-(I_m-\bar{I}_{g(e)})^2/(2D)]  \nl
&\times& \exp[-(Q_m-\bar{Q}_{g(e)})^2/(2D)] \;,
\end{eqnarray}
with $D=1/t_m$ the same as in the single-quadrature measurement.
Respectively,
the average of each quadrature output is given as
\bea
\bar{I}_{g(e)} = \frac{1}{t_m}\int^{t_m}_{0}\!\!\!dt \; \bar{I}_{g(e)}(t)
\eea
with $\bar{I}_{g(e)}(t)=\sqrt{2\kappa}~{\rm Re}[\alpha_{g(e)}(t)]$;
and
\bea
\bar{Q}_{g(e)} = \frac{1}{t_m}\int^{t_m}_{0}\!\!\!dt \; \bar{Q}_{g(e)}(t)
\eea
with $\bar{Q}_{g(e)}(t)=\sqrt{2\kappa}~{\rm Im}[\alpha_{g(e)}(t)]$.
As pointed out previously, tuning $\Delta_r=0$ and
with an initial vacuum cavity state,
the imaginary parts of $\alpha_{g}(t)$ and $\alpha_{e}(t)$ are equal.
This means that the $Q$-quadrature output $Q_m$ does not provide
qubit state information. However, in an ideal case,
tracking this output simultaneously with the $I$-quadrature outcome ($I_m$)
can maintain the joint state of the qubit-plus-cavity in a pure state.
So, with respect to the optimal single-quadrature measurement ($\varphi=0$),
the $(I,Q)$ two-quadrature measurement reduces the signal
$|\bar{I}_e(t_m)-\bar{I}_g(t_m)|$ by a factor of $1/\sqrt{2}$.

With the knowledge of $P_{g(e)}(I_m,Q_m)$,
the diagonal elements of the qubit state, $\rho_{gg}$ and $\rho_{ee}$,
can be determined straightforwardly
using the Bayesian rule \Eq{Bayes3a-1}
for their informational
evolution conditioned on the outcome $(I_m,Q_m)$.
For the off-diagonal elements, the quantum Bayesian rule is
the same as \Eq{Bayes3b-1}, with the phase factor $\Phi_1(t_m)$
unchanged but $\Phi_2(t_m)$ now given by
\begin{eqnarray}\label{Phi2a}
\Phi_2(t_m) = -\int_0^{t_m}\!\!\! \sqrt{\Gamma_{m}(t)/2}\;
\widetilde{Q}_m(t)\,dt \,.
\end{eqnarray}
$\widetilde{Q}_m(t)$ can be determined as follows. 
For the second branch $Q$-quadrature measurement,
we have $Q_m(t)= I_{\varphi}(t)|_{\varphi=\pi/2}
= \sqrt{\kappa}|\mu(t)|\sin\theta_{\mu} +\xi_2(t)$,  
by noting that $\varphi=\pi/2$ and $\Gamma^{(2)}_{ci}(t)=0$. 
Similar to the consideration leading to \Eq{Phi2a-1}, 
we determine then
$\widetilde{Q}_m(t)= Q_m(t)-\sqrt{\kappa}|\mu(t)|\sin\theta_{\mu}$ 
in \Eq{Phi2a} for $\Phi_2$.

\begin{figure}
  \includegraphics[scale=0.9]{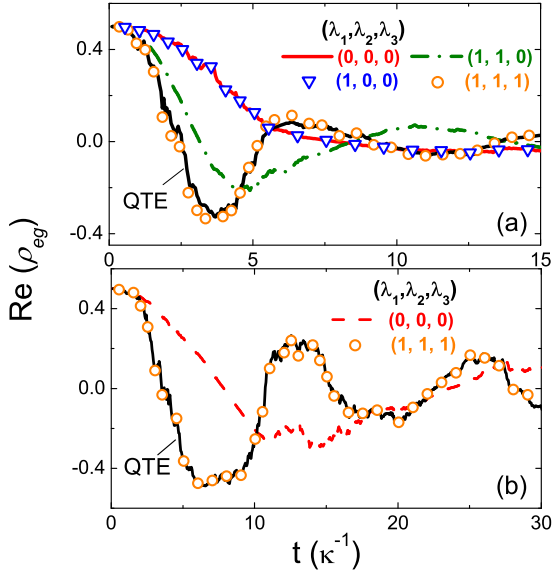}
\caption{ (Color online)
Correcting effects to the ``bare" Bayesian rule
and comparison with the (exact) QTE results
(off-diagonal element of the qubit state),
summarized in a unified form:
$\rho_{ge}(t)= \tilde{\rho}_{ge}(t)
|\langle \alpha_{e}(t)| \alpha_{g}(t)\rangle|^{\lambda_1}
e^{-i[\lambda_2 \Phi_1(t)+\lambda_3  \Phi_2(t)]}$,
with $\lambda_{1,2,3}=0$ or $1$,
and $\tilde{\rho}_{ge}(t)$, $\Phi_1(t)$
and $\Phi_2(t)$ as defined in the main text.
The qubit is assumed with an initial state
$(|e\rangle+|g\rangle)/\sqrt{2}$.
Panel (a) shows the result corresponding to
the single-quadrature measurement
with local oscillator phase $\varphi=\pi/4$,
while the result for two quadratures is shown in (b).
Parameters in each case are:
$\Delta_r=0$, $\chi=0.1$, $\epsilon_m=1.0$, and $\kappa=2.0$.
}\label{Bayesian}
\end{figure}

\section{Numerical Results and Discussions}

\subsection{Effects of Corrections}

We show the correction effects in Fig.\ 2
and demonstrate the proposed Bayesian rule
by comparison with the exact results from simulation of \Eq{QTE},
in (a) for single-quadrature and (b) for two-quadrature measurements.
For the case of the single-quadrature measurement,
we have used \Eq{Phi2} to approximate $\Phi_2(t_m)$
for the purpose of revealing the quality of the rule
using only the integrated quadrature.
For both types of measurements,
we find that the proposed quantum Bayesian rule
can give reliable estimate for the qubit state.

%


In Fig. 2(a) each correction is presented individually
for the single-quadrature measurement.
Since the individual effect of each correction term is similar,
only the total result is shown for the
two-quadrature case illustrated in Fig. 2(b).
Whereas the consequences of the phase factors are dramatic,
one may notice that in this plot the correction from the
purity-degradation-factor is very weak (almost negligible).
However, its physical meaning is clear.
It characterizes the {\it intrinsic} purity of the qubit
state imposed by the cQED measurement in the ideal case.
Actually, the qubit-state purity associated with Fig.\ 2
is about 0.97, but not unity. Changing the parameters,
one can make this correction effect more prominent,
as to be shown in Sec.\ VI C.


\subsection{Limiting Cases}

We now consider the three correction factors in limiting cases,
making in particular a connection with the work by Korotkov \cite{Kor11}.
First, for the purity degradation factor
$|\langle \alpha_{e}(t_m)| \alpha_{g}(t_m)\rangle|$,
based on the solution in Appendix A
we obtain, in steady state:
\bea
D=|\langle \bar{\alpha}_{e}| \bar{\alpha}_{g}\rangle|
= \exp\left[ -\frac{2\epsilon^2_m\chi^2}{(\chi^2+\kappa^2/4)^2}  \right] \,.
\eea
Further, in the bad-cavity and weak-response limit,
$\kappa\gg\chi$, the result simplifies to
$|\langle \bar{\alpha}_{e}| \bar{\alpha}_{g}\rangle|
= \exp[ -8\bar{n}(\chi/\kappa)^2]$,
where $\bar{n}=|\alpha_0|^2$
with $\alpha_0=-i\epsilon_m/(\frac{\kappa}{2})$.
Accordingly, only in the limit $\kappa\gg\chi$
and with small $\bar{n}$ (cavity photon number),
can the purity factor be approximated to unity,
implying a pure state of qubit
under the quadrature measurement.
Otherwise, this factor should be taken into account.
In particular, this implies that for the not very bad cavity
and with $\bar{n}$ not very small,
the $D$ factor cannot be treated as unity.
For instance, as to be seen below in Fig.\ 4(a),
the $D$ factor will reduce to a value lower than 0.8
for $\chi=0.1\kappa$ and $\bar{n}\simeq 4$.
We note that this factor was not addressed in Ref.\ \cite{Kor11}.
But in a recent report \cite{Sid14}
the similar $D$ factors were included
in the {\it concurrence} calculation for a two-qubit cQED system
(see Eq.\ (12) in the Supplementary Information of Ref.\ \cite{Sid14}).   

Let us now consider $\Phi_1(t_m)$.
The key quantity associated
is the effective magnetic field
$B(t)=2\chi{\rm Re}[\alpha_g(t)\alpha^*_e(t)]$.
Physically speaking, it describes a generalized
(time dependent) ac-Stark effect.
From the analytic solution in Appendix A,
we formally reexpress
$\alpha_{g(e)}(t)=\mp a(t)+ib(t)$, which leads to
$\alpha_{g}\alpha^*_{e}=-(a^2-b^2)+2 i ab$.
We see that the imaginary part ($2ab$)
determines the dephasing rate $\Gamma_d$,
and the real part ($b^2-a^2$) affects the qubit energy.
Further, in steady state, we have
\bea
B=\frac{2\chi\epsilon^2_m}{d^2+\kappa^2\chi^2/d^2} \,,
\eea
where $d^2\equiv \Delta^2_r-\chi^2+\kappa^2/4$.
Again, in the bad-cavity and weak-response limit,
the result reduces to $B\simeq 2\chi\bar{n}$.
This is the standard ac-Stark shift to the qubit energy.

Finally, let us consider $\Phi_2(t_m)$.
Based on the steady-state solution of the  cavity fields,
we obtain the back-action rate as
\bea
\Gamma_{ba}=\kappa\frac{4\epsilon^2_m\chi^2}
{(\chi^2+\kappa^2/4)^2}\sin^2\varphi \,.
\eea
Moreover, in the limit $\kappa\gg\chi$, the result is further simplified as
$\Gamma_{ba}\simeq 16\bar{n}\kappa(\chi/\kappa)^2\sin^2\varphi$.
Substituting this result into the expression of $\Phi_2(t_m)$,
we find that we recover the ``realistic"-back-action induced
phase factor in the bad-cavity and weak-response limit,
obtained in Ref.\ \cite{Kor11}
by using photon-caused qubit rotation considerations.
We note also that, in the context of $(I,Q)$
two-quadrature measurements, the state-update rule
constructed in Ref.\ \cite{Dev13}
(in the Supplementary Materials)
contains as well this same factor in the same limiting case.
However, while our approach can derive theirs, it seems
to be an open problem how to use their approaches
to derive some results here.

\subsection{ Non-Limiting Cases }

In this subsection we present some numerical results
beyond the ``bad-cavity" and weak-response limits,
and compare with the Bayesian rule constructed in Ref.\ \cite{Kor11}.
Cases that violate the restrictive limits can be the following:
(i) the quality factor of the cavity is relatively high,
as required in quantum information processing
in order to employ the cavity photon as a data-bus;
(ii) the qubit-cavity coupling ($\chi$ in the dispersive regime) is strong,
which is required for quantum information processing
and would violate the weak-response assumption;
and (iii) the average cavity photon number $\bar{n}$ is not tiny,
which has the advantage of enhancing the measurement signal to overcome
the noise from the amplifiers and circuits.

For the sake of brevity, we denote the Bayesian rule
proposed in Ref.\ \cite{Kor11} by BR-${\rm I}$,
the rule constructed in the present work by BR-${\rm II}$,
and the one involving
the approximate $\Phi_2$ of \Eq{Phi2} by BR-${\rm II'}$.
In Fig.\ 3 we display results for both the single-
and two-quadrature measurements
outside the ``bad-cavity" and weak-response limits.
Clearly, we find that in this case
the purity-degradation factor,
$D(t)=|\la \alpha_e(t)|\alpha_g(t) \ra|$,
is reduced to values obviously lower than unity,
and for both measurements
BR-${\rm II}$ fits the (exact) QTE results
(off-diagonal element of the qubit state)
better than BR-${\rm I}$.
For single-quadrature measurement in this non-limiting case,
we find the use of the
approximate phase factor $\Phi_2$ causes some
deviation from the precise one, as revealed in Fig.\ 3(c).
However, combining with the other two corrections
($D$ and $\Phi_1$), BR-${\rm II}'$
can give reasonable results as shown in Fig.\ 3(e).

\begin{figure}
  \includegraphics[scale=0.75]{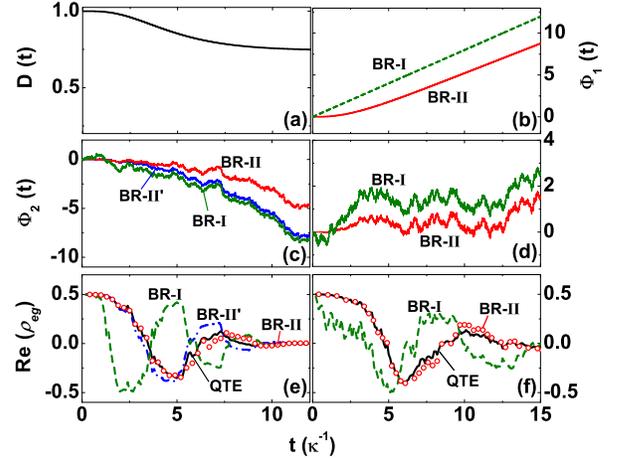}
\caption{ (Color online)
Comparison of the Bayesian rules gainst the exact QTE
in non-limiting case,
for both single-quadrature (with $\varphi=\pi/4$)
and two-quadrature measurements.
In (a) and (b) we plot the common purity-degradation factor
$D(t)=|\la \alpha_e(t)|\alpha_g(t) \ra|$
and phase factor $\Phi_1$,
while the results in (c) and (e) correspond to
single-quadrature measurement,
and the results in (d) and (f) to two-quadrature measurement.
BR-${\rm I}$ and BR-${\rm II}$ denote
the Bayesian rules proposed in Ref.\ \cite{Kor11}
and in the present work,
whereas BR-${\rm II'}$ involves the approximation of \Eq{Phi2}.
We assume the qubit in initial state
$(|e\rangle+|g\rangle)/\sqrt{2}$
and the setup parameters
$\Delta_r=0$, $\chi=0.1\kappa$ and $\epsilon_m=\kappa$.
These parameters indicate the cavity photon number
$\bar{n}\simeq 4$.   }
\end{figure}

\subsection{ Experimental Issues }

In order to implement the Bayesian rule proposed
in Sec.\ IV and V in experiments,
the key quantities to be fixed are the cavity fields
$\alpha_g(t)$ and $\alpha_e(t)$.
Viewing the analytic solutions in Appendix A, the cavity
damping rate $\kappa$ and the dispersive coupling $\chi$
should be determined in advance.
For a given detuning $\Delta_r$ and driving amplitude $\epsilon_m$,
these two parameters can be extracted from the steady-state
mean values of the quadrature measurements,
$\bar{I}_{g}$ and $\bar{I}_{e}$, which are related
to $\bar{\alpha}_{g}$ and $\bar{\alpha}_{e}$.
With these extracted parameters at hand, one can
accordingly implement the proposed Bayesian rule,
with the cavity in an initial vacuum
or certain known steady state (e.g., $|\bar{\alpha}_g\ra$).

In experiments, one needs also to properly account
for the unavoidable measurement inefficiencies,
in particular the circuit and amplifier noises.
This important issue has been addressed
by Wiseman {\it et al.} in a series of papers
\cite{noise-1,noise-2,noise-3}, where the so-called
realistic quantum trajectory equation has been developed.
However, the resultant equation would be difficult
to use in practice.
Within the framework of the Baysian approach,
it seems simpler to address this issue \cite{noise-4},
since accounting for the extra noise only
corresponds to a more Bayesian inferring.
As a result,
the effect of the extra noise requires us to partially
average the ideal-measurement-result conditioned state.
We will incorporate our present quantum Bayesian approach
with this type of treatment in a separate work.

Finally, we mention that so far there have been a few
experiments involving the quantum Bayesian rule
in the bad-cavity and weak-response regime \cite{Sid12,Dev13,Sid13}.
In the feedback control experiment \cite{Sid12},
the phase-sensitive detection scheme corresponds to a case
where the phase factor $\Phi_2(t_m)$ in \Eq{Bayes3b-1} vanishes.
Moreover, in the bad-cavity and weak-response limit,
the factor $\Phi_1$ reduces to $2\chi\bar{n}t_m$
and the purity degradation factor is about unity.
In two recent experiments \cite{Dev13,Sid13}, however,
the phase factor $e^{-i\Phi_2(t_m)}$ is present
and was demonstrated with satisfactory accuracies.
In view of these remarkable advances in cQED measurements,
further experiments to demonstrate the quantum Bayesian rule
in more general cases would be of great interest.

\section{Concluding Remarks}

To summarize, we have constructed a quantum Bayesian rule
for weak measurements of qubits in cQED.
Our construction was guided by a microscopic analysis of the cavity field
and the cavity-photon-eliminated effective QTE.
This type of treatment provided a different route from
Korotkov's method in Ref.\ \cite{Kor11}.
The present work, for both the single- and two-quadrature measurements,
generalizes the results in Ref.\ \cite{Kor11}
from ``bad-cavity" and weak-response limits to more general conditions.
Numerical comparisons with the direct QTE simulations show
that the proposed rule can work with high accuracy
even in non-limiting cases.


We would like to remark that, originally, the Bayesian approach
was not constructed or integrated from QTE \cite{Kor99,Kor11}.
Their mathematical connection is also not very straightforward.
For infinitesimal short-time evolution, their equivalence
can be proved. However, for longer time state update,
there exist slight numerical differences, despite
the reasonable agreements observed in Figs.\ 2 and 3.
Rather than an exact derivation from QTE,
we would like to regard the Bayesian rule as a construction.
In particular, the cavity-photon-eliminated effective QTE
contains unusual {\it stochastic} unitary term.
As an {\it ansatz}, we inserted (integrated) it into the Bayesian dynamics
of the off-diagonal elements and revealed an interesting
connection with the results of Ref.\ \cite{Kor11} in limiting cases.

It has become clear that,
in addition to the quantum trajectory equation,
the empirical Bayesian formulas are
very useful in experiments \cite{Dev13,Sid13,Sid12}.
In connection with the experiment of Ref.\ \cite{Dev13},
a POVM-type formalism was constructed for qubit state
update in cQED, which is actually equivalent to the result
obtained in Ref.\ \cite{Kor11}. Also, both results work
in the ``bad-cavity" and weak response limits.
Their minor difference is:
the (POVM) measurement operator (i.e., $M_{I_m,Q_m}$
in Eq.\ (6) in the Supplementary Materials of Ref.\ \cite{Dev13}),
avoids the ``purity" consideration in constructing the Bayesian rule
and contains the phase factors discussed
in Ref.\ \cite{Kor11} and in our present work.
Finally, at the revision stage of this work,
we became aware of the new publication \cite{Sid14}, in which similar
{\it purity degradation factor} was included
in the concurrence calculation for a two-qubit cQED system
(see Eq.\ (12) in the Supplementary Information).
However, it seems that in this work the Bayesian rule
proposed in Ref.\ \cite{Kor11}, particularly the
``realistic" back-action phase factor
($\Phi_2$ denoted in our present work) was not involved.
There, the quantum trajectory equation approach
was also used to compare with the experimental quantum trajectories,
see, Eqs.\ (14) and (22) in the Supplementary Information.
Viewing these remarkable efforts and progress,
we expect our proposed Bayesian rule to be useful,
for both theoretical interests and
future cQED measurement and control experiments.       

\vspace{0.8cm}
{\it Acknowledgments.}---
The authors thank Justin Dressel for valuable discussions
and acknowledge the support and hospitality from RIKEN
during their visit.
This work is supported by the NNSF of China and
the Major State Basic Research
Project under Grant Nos.\ 2011CB808502 \& 2012CB932704.

\appendix

\section{Cavity Fields and Measurement Principle}

The interplay of external driving and cavity damping would evolve
the cavity field as an optical coherent state, with the time-dependent
coherence parameter determined by
\begin{eqnarray}\label{coh-state}
 && \dot{\alpha_g}(t)=-i\epsilon_m
 -i(\Delta_r-\chi)\alpha_g(t)-\kappa\alpha_g(t)/2 \,,   \nonumber\\
 && \dot{\alpha_e}(t)=-i\epsilon_m
 -i(\Delta_r+\chi)\alpha_e(t)-\kappa\alpha_e(t)/2  \,. \nl
\end{eqnarray}
Here, corresponding to the qubit state $|g(e)\rangle$,
the frequency of the cavity has a dispersive shift $\mp\chi$.
Moreover, analytically solving these two equations yields
\begin{eqnarray}
&& \alpha_g(t)=\bar{\alpha}_g \big[1-e^{-i\Delta^{(-)}_r t-\kappa t/2}\big]
+\alpha_0 e^{-i\Delta^{(-)}_rt-\kappa t/2} \,, \nonumber\\
&& \alpha_e(t)=\bar{\alpha}_e \big[1-e^{-i\Delta^{(+)}_rt-\kappa t/2}\big]
+\alpha_0 e^{-i\Delta^{(+)}_rt-\kappa t/2} \,. \nl
\end{eqnarray}
For simplicity we introduce
$\Delta^{(\pm)}_r=\Delta_r\pm\chi$.
In this solution, $\alpha_0$ is the initial cavity field
before the dispersive measurement, which is zero
if one starts the measurement with a vacuum state. Also,
$\bar{\alpha}_{g(e)}=-i\epsilon_m/[i\Delta^{(\mp)}_r +\kappa/2]$,
are the respective steady-state fields.

Based on the above solutions, one can understand the basic principle
of quadrature measurement which is employed to extract the information
of the qubit state. For a single-quadrature homodyne measurement,
the observable (measurement operator) is
$\hat{I}_{\varphi}=\frac{1}{2}(ae^{-i\varphi}+a^{\dagger}e^{i\varphi})$,
with $\varphi$ the local oscillator (LO) phase.
Corresponding to the qubit state $|g(e)\ra$,
the average quadrature output is
\bea
\bar{I}_{g(e)}(t)=2\sqrt{\kappa}\,{\rm Re}[\alpha_{g(e)}(t)e^{-i\varphi}]\,.
\eea
Denoting
\bea
\alpha_{e}(t)-\alpha_{g}(t)=|\beta(t)|e^{i\theta_{\beta}} \,,
\eea
we have
\bea
\bar{I}_{e}(t)-\bar{I}_{g}(t)
= 2\sqrt{\kappa}|\beta(t)|\cos(\theta_{\beta}-\varphi) \,.
\eea
From this result, it becomes clear that the
optimal qubit-state information can be inferred
if we tune the LO phase to $\varphi=\theta_{\beta}$.
In contrast, if we tune $\varphi=\theta_{\beta}+\pi/2$,
the quadrature measurement does not
reveal any information for the qubit-state \cite{Wis12}.
More specifically, starting the measurement with an empty cavity
and choosing a resonant measurement driving
($\Delta_r=\omega_r-\omega_m=0$), one can easily prove that
$\theta_{\beta}=0$.
This means that, choosing $\varphi=0$, one can achieve maximal
information for the qubit state, while no qubit state
information can be inferred if choosing $\varphi=\pi/2$.

Moreover, for a single realization of quadrature measurement
during $(0,t_m)$, we introduce the mean integrated quadrature output
$I_m = \frac{1}{t_m}\int^{t_m}_{0}dt I_{\varphi}(t)$,
where the continuous outcome is described by
$I_{\varphi}(t)=\sqrt{\kappa}\la a e^{-i\varphi}
+a^{\dagger}e^{-i\varphi}\ra_{\varrho(t)} + \xi(t)$.
Corresponding to qubit state $\ket{g(e)}$, the distribution
$P_{g(e)}(I_m)$ of the integrated quadrature $I_m$
is Gaussian, $P_{g(e)}(I_m)=\frac{1}{\sqrt{2\pi D}}
\exp[-(I_m-\bar{I}_{g(e)})^2/(2D)]$.
Here, $\bar{I}_{g(e)}$ is the average quadrature outcome, given by
$\bar{I}_{g(e)} = \frac{1}{t_m}\int^{t_m}_{0}dt \bar{I}_{g(e)}(t)$.
The variance $D$ can be analytically determined as follows.
Consider the expression of the homodyne current.
Noting that the first term describes the average current,
the deviation of the individual quadrature output ($I_m$)
from the averaged $\bar{I}_{g(e)}$ is thus caused
by the second term $\xi(t)$.
Denoting $\tilde{I}(t)\equiv \xi(t)$, from definition we have
\bea
\la \tilde{I}^2\ra = \int^{\infty}_{-\infty} d\tilde{I}
 \, P(\tilde{I}) \, \tilde{I}^2 = D \;.
\eea
On the other hand, via a direct calculation,
\bea
\la \tilde{I}^2\ra
= \frac{1}{t_m^2} \int^{t_m}_{0}\!\!dt_1 \int^{t_m}_{0}\!\!dt_2
\la \xi(t_1) \xi(t_2) \ra = \frac{1}{t_m} \;,
\eea
we thus obtain $D=1/t_m$.

\section{Revisit the Purity Degradation Factor}

Based on the transformed \Eq{pQTE}, in this Appendix
we revisit the purity degradation factor
$|\la \alpha_e(t_m)|\alpha_g(t_m)\ra |$.
First, we notice that the last term of \Eq{pQTE} does not
only play the usual role of unitary evolution,
but also has an effect of unraveling the qubit state.
This can be clearly seen by setting $\varphi=\pi/2$.
In this case, $\Gamma_{ci}=0$,
then one may expect that the r.h.s. second Lindblad term
will gradually completely dephase the qubit state.
However, interestingly, the last term will prevent this,
owing to its certain unraveling ability.
This feature is in agreement with the Bayesian rule
for the case of identical $P_g(I_m)$ and $P_e(I_m)$.
Both approaches predict that the qubit will be maintained
in a superposition state with high purity.
Even better, we can combine the last two terms
into a single unraveling term:
$\sqrt{\Gamma_m(t)}/2{\cal H}[\Lambda\sigma_z]\rho(t)\xi(t)$,
where $\Gamma_m(t)=\Gamma_{ci}(t)+\Gamma_{ba}(t)=\kappa|\beta(t)|^2$
(the measurement rate) is independent of the choice of $\varphi$,
and $\Lambda=\cos(\varphi-\theta_{\beta})-i\sin(\varphi-\theta_{\beta})$
depends on $\varphi$.
We may interpret this result as follows:
$\Gamma_m$ determines the unraveling extent;
and $\Lambda\sigma_z$ describes the unraveling (measuring) means.
We propose then a dephasing factor of the form
$\exp\{-\int_0^{t_m} dt [\Gamma_d(t)-\Gamma_m(t)/2]\}$.

Below, we prove that the two expressions
of the purity degradation factor
for the qubit state are identical, i.e.,
\bea
\exp\left\{-\int_0^t \!\!dt'[\Gamma_d(t')-\Gamma_m(t')/2]\right\}
=|\langle\alpha_{g}(t)|\alpha_{e}(t)\rangle| \,.  \nonumber\\
\eea
This is equivalent to proving the following:
\bea\label{eq-dec}
&& -\int_0^t dt'\big[\Gamma_d(t')-\Gamma_m(t')/2\big]   \nl
&=&  -\frac{1}{2} [|\alpha_{e}(t)|^2+|\alpha_{g}(t)|^2]
+\Re[\alpha_{e}(t)\alpha_{g}^*(t)] \,. \nonumber\\
\eea
In obtaining this result,
the property of coherent states has been used.
Under the conditions $\Delta_r=0$ and $\alpha_{e}(0)=\alpha_{g}(0)=0$,
more explicitly we reexpress the solution of \Eq{coh-state} as
 \begin{eqnarray}\label{solution2}
 && \alpha_e(t)=\frac{i\epsilon_m}{i\chi+\kappa/2}
 \left[ e^{-(i\chi+\kappa/2)t} -1  \right]  \,,           \nonumber\\
 && \alpha_g(t)=\frac{i\epsilon_m}{-i\chi+\kappa/2}
 \left[ e^{(i\chi-\kappa/2)t} -1 \right]     \,.         \nonumber\\
 \end{eqnarray}
Substituting these two expressions into the l.h.s of \Eq{eq-dec} gives
  \begin{eqnarray}
 &&  -\int_0^t dt' \big[ \Gamma_d(t')-\Gamma_m(t')/2 \big]        \nonumber\\
 &=& -2\chi\int_0^t dt' \Im \big[\alpha_{g}(t')\alpha_{e}^*(t')\big] \nonumber\\
 && +\frac{\kappa}{2}\int_0^t dt' |\alpha_{e}(t')-\alpha_{g}(t')|^2    \nonumber\\
 &=& \frac{\kappa}{2}\int_0^t dt' \big[\alpha_{e}(t')\alpha_{e}^*(t')
 +\alpha_{g}(t')\alpha_{g}^*(t')\big]   \nonumber\\
 && +(i\chi-\kappa/2)\int_0^t dt' \alpha_{e}^*(t')\alpha_{g}(t')  \nonumber\\
 && -(i\chi+\kappa/2)\int_0^t dt' \alpha_{e}(t')\alpha_{g}^*(t') \,.
\end{eqnarray}
Further evaluation yields:
\begin{eqnarray}
 && -\int_0^t dt' \big[\Gamma_d(t')-\Gamma_m(t')/2 \big]        \nonumber\\
 &=& -\frac{\epsilon_m^2}{\chi^2+\kappa^2/4}\big[e^{-\kappa t}+1
 -e^{(i\chi-\kappa/2)t}-e^{(-i\chi-\kappa/2)t}\big]  \nonumber\\
 && +\frac{\epsilon_m^2}{2(i\chi-\kappa/2)^2}
 \big[e^{2(i\chi-\kappa/2)t}+1-2e^{(i\chi-\kappa/2)t}\big] \nonumber\\
 && +\frac{\epsilon_m^2}{2(-i\chi-\kappa/2)^2}
 \big[e^{2(-i\chi-\kappa/2)t}+1-2e^{(-i\chi-\kappa/2)t}\big]  \nonumber\\
 &=& -\frac{1}{2}(|\alpha_{e}(t)|^2+|\alpha_{g}(t)|^2)
 +\Re[\alpha_{e}(t)\alpha_{g}^*(t)] \,,
\end{eqnarray}
showing the validity of Eq.\ (B2) and thus of Eq.\ (B1).


\end{document}